\documentclass[traditabstract, longauth]{aa}  
\usepackage{graphicx}
\usepackage{txfonts}
\usepackage{natbib}
\bibpunct{(}{)}{;}{a}{}{,}

\newcommand{\eh}[1]{\,\mathrm{#1}}

\newcommand{\gev}{\eh{GeV}}

\newcommand{\dg}{^{\circ}}

\newcommand{\pct}{\eh{\%}}

\newcommand{\pcnt}{$\pct$}

\newcommand{\mr}[1]{\mathrm{#1}}

\renewcommand{\epsilon}{\varepsilon}

\newcommand{\tin}[1]{_{\mr{#1}}}

\newcommand{\bref}[1]{#1}
\newcommand{\breftwo}[1]{#1}
\newcommand{\brefthree}[1]{#1}

\begin{document}
   \title{Phase-resolved energy spectra of the Crab pulsar in the
range of $50$-$400\gev$ measured with the MAGIC telescopes}


%
\author{
 J.~Aleksi\'c\inst{1} \and
 E.~A.~Alvarez\inst{2} \and
 L.~A.~Antonelli\inst{3} \and
 P.~Antoranz\inst{4} \and
 M.~Asensio\inst{2} \and
 M.~Backes\inst{5} \and
 J.~A.~Barrio\inst{2} \and
 D.~Bastieri\inst{6} \and
 J.~Becerra Gonz\'alez\inst{7,}\inst{8} \and
 W.~Bednarek\inst{9}
% ...
 \newline
 \textit{$^{*}$Please address correspondence to:}\, klepser@ifae.es, giavitto@gmail.com,
tysaito@mpp.mpg.de\bref{, hirotani@tiara.sinica.edu.tw}}

\institute { IFAE, Edifici Cn., Campus UAB, E-08193 Bellaterra, Spain
 \and Universidad Complutense, E-28040 Madrid, Spain
 \and INAF National Institute for Astrophysics, I-00136 Rome, Italy
 \and Universit\`a  di Siena, and INFN Pisa, I-53100 Siena, Italy
 \and Technische Universit\"at Dortmund, D-44221 Dortmund, Germany
 \and Universit\`a di Padova and INFN, I-35131 Padova, Italy
 \and Inst. de Astrof\'{\i}sica de Canarias, E-38200 La Laguna, Tenerife, Spain
 \and Depto. de Astrof\'{\i}sica, Universidad de La Laguna, E-38206 La Laguna, Spain
 \and University of \L\'od\'z, PL-90236 Lodz, Poland
% ...
}

   \date{Accepted version, 16.02.2012}

  \abstract
   {We use $73\eh{h}$ of stereoscopic data taken with the MAGIC telescopes to
investigate the very high-energy (VHE) gamma-ray emission of the Crab pulsar.
Our data show a highly significant pulsed signal in
the energy range from $50$ to $400\gev$ in both the main pulse (P1) and the
interpulse (P2) phase regions.
We provide the widest
spectra to date of the VHE components of both peaks, and these spectra extend
to the
energy range of satellite-borne observatories.
The \bref{good resolution and background rejection of the stereoscopic MAGIC
system} allows us to cross-check the
correctness of each spectral point of the pulsar by comparison with the
corresponding (strong and well-known) Crab nebula flux.
The spectra of both P1 and P2 are compatible with
power laws with photon indices of $4.0\pm0.8$ (P1) and $3.42\pm0.26$ (P2),
respectively, and the
ratio P1/P2 between the \brefthree{photon counts of the two} pulses is
$0.54\pm0.12$. The VHE emission can be understood
as an additional component produced by the inverse Compton scattering of
secondary and tertiary $e^{\pm}$ pairs on IR-UV photons.
}


   \authorrunning{Aleksi\'c et al.}
   \titlerunning{Phase-resolved Crab pulsar spectra with MAGIC}
   \maketitle
%

\section{Introduction}

The Crab pulsar is a young neutron star that is the central remnant of the
supernova SN 1054 \citep{sn1054}. It is one of the few pulsars that have been detected in
almost all energies, ranging from radio \citep[e.g.,][]{jodrellbank} to
VHE gamma rays. In the highest-energy regime, it was
detected up to a few tens of GeV by Fermi-LAT \citep{fermicrab}, between
approximately $25-100\eh{GeV}$ by
MAGIC \citep{magicsciencecrab, takathesis, magictakacrab} and above $100\gev$
by VERITAS \citep{veritascrab}. The light curves and the spectra obtained by these
observations suggest that gamma-ray pulsars have high-altitude emission
zones that avoid a super-exponential spectral cutoff, which would be caused by magnetic pair
\brefthree{production.}
Consequently, the favored models \bref{to explain the production of gamma
rays to at least a few GeV} are \brefthree{those in which fan-like beams of
high-energy electrons scan over a large fraction of the outer magnetosphere,
either}
very close to the light cylinder \citep[outer gap model,][]{outergap, outergap2}
or all along the last open field lines \citep[slot gap model,][]{slotgap, slotgap2}.

For other rotation-powered gamma-ray pulsars beside the Crab pulsar,
Fermi-LAT observations have shown that
their energy spectra exhibit exponential cutoffs
at around a few GeV \citep{fermipulsars}.
This mild cutoff has been widely accepted as a result of
the curvature process by $e^{\pm}$ migrating along curved paths. In this
scenario, the cutoff energy corresponds to the
highest characteristic curvature-radiation energy
of the particles accelerated in the magnetosphere
\citep[e.g.,][]{outergap2}.
However, the spectrum of the Crab pulsar
strongly disfavors an exponential cutoff \citep{magictakacrab, veritascrab},
making this pulsar a counterexample of the general property.
Thus, to develop pulsar emission theories
beyond the widely accepted curvature-radiation models,
it is essential to examine the detailed phase-resolved spectrum
of this youngest pulsar in the HE to VHE regimes.

\section{Data set and analysis techniques}

The two MAGIC telescopes~\citep{magicstereoperformance, icrccrabnebula}
situated on the island of La Palma (28.8\degr~N, 17.8\degr~W,
$2220\eh{m\,a.s.l.}$), use the imaging atmospheric Cherenkov
technique
to detect gamma rays above a few tens of GeV\footnote{\bref{The nominal} threshold in standard
trigger mode, defined as the peak of the simulated energy distribution for a
Crab-nebula-like spectrum after
all cuts and at low zenith angles, is $75-80\gev$.}. Since summer 2009, when
the system started operating in stereoscopic mode,
its background suppression was substantially improved, and a
sensitivity\footnote{Defined as the source strength needed to achieve $N\tin{ex}/\sqrt{N\tin{bkg}}=5$ in $50\eh{h}$ effective
on-time.}  of
$0.8\pct$ Crab nebula units above $250\gev$ \breftwo{has been} achieved
\citep{magicstereoperformance}. 

In the analysis presented here, we used $73\eh{h}$ of good quality stereoscopic data from the winter seasons
in 2009/2010 and 2010/2011. Of these data, $43\eh{h}$ were taken in the wobble
observation mode \citep{wobble}, and another $30\eh{h}$ in
on-source observation mode.
The data were taken
at zenith angles below $35\dg$ to ensure a low threshold.

For the data analysis, we used the standard MAGIC analysis package MARS
\citep{mars, magicstereoperformance}, applying  the so-called \textit{sum} cleaning
\citep{icrcadvanced} to achieve the lowest possible threshold.
For the gamma/hadron separation and gamma direction
estimation we apply the random forest (RF) technique \citep{rf}. Because our
background is dominated by
Crab nebula gamma rays instead of hadrons
already above $\sim120\gev$, we opted for loose and conservative
selection cuts.
The phase of each event with respect to the main radio pulse was calculated
using the TEMPO2 package \citep{tempo2} and the monthly ephemerides publicly provided by
the Jodrell Bank
Observatory\footnote{http://www.jb.man.ac.uk/research/pulsar/crab.html} \citep{jodrellbank}.

%

For the spectra, we applied the unfolding algorithms described in
\citet{unfolding}. \bref{This procedure corrects the migrations and the energy biases
expected in the threshold regime. During unfolding iterations, the simulated events are reweighted each time with the appropriate
spectrum derived in the previous iteration.}

\section{Results}

\subsection{\breftwo{Folded} light curves}

   \begin{figure}
   \centering
   \includegraphics[width=9.5cm]{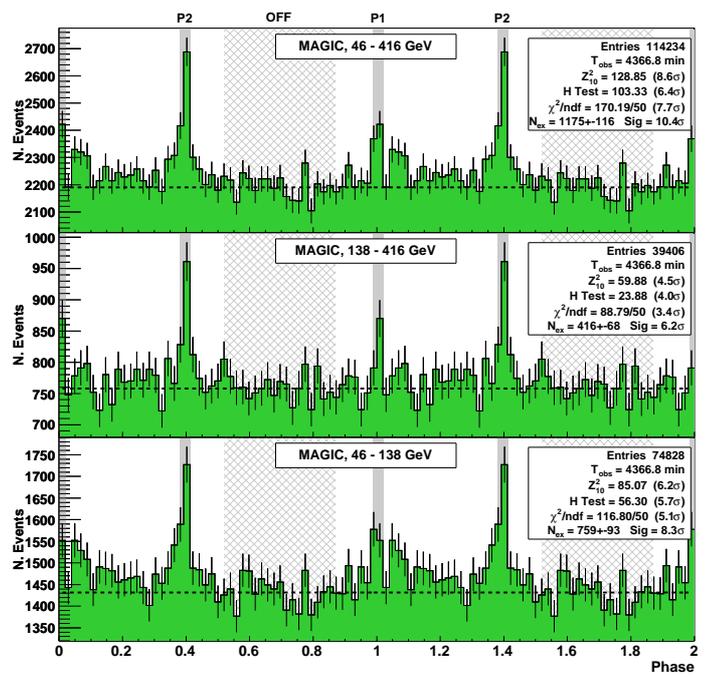}
      \caption{MAGIC folded light curves of the Crab pulsar for our total range in
estimated energy and for two separate sub-bins. The shaded areas are the on-phase regions
P1$_\mr{M}$ and P2$_\mr{M}$ (see text), the light shaded area is the off-region $[0.52 -
0.87]$. The dashed line is the constant background level calculated from that
off-region.
              }
         \label{FigLC}
   \end{figure}

We obtained three folded light curves using all data with estimated energies
between $46-416\gev$
and for two sub-ranges  $46-138\gev$ and
$138-416\gev$ (Fig.\ref{FigLC}). The median true energies of these samples
were estimated from simulations to be
approximately $100\gev$, $80\gev$ and $180\gev$, respectively.
The significance of the pulsation was tested with the $Z^2_{10}$ test,
the H test \citep{htest}, and a simple $\chi^2$-test. None of these tests
makes an a priori assumption concerning the position and the shape of the
pulsed emission, and they yield significances of $8.6\eh{\sigma}$,
$6.4\eh{\sigma}$ and $7.7\eh{\sigma}$, respectively. The folded light curve clearly
shows two distinct peaks, the well-known P1 and P2.

   \begin{figure}
   \centering
   \includegraphics[width=8.5cm]{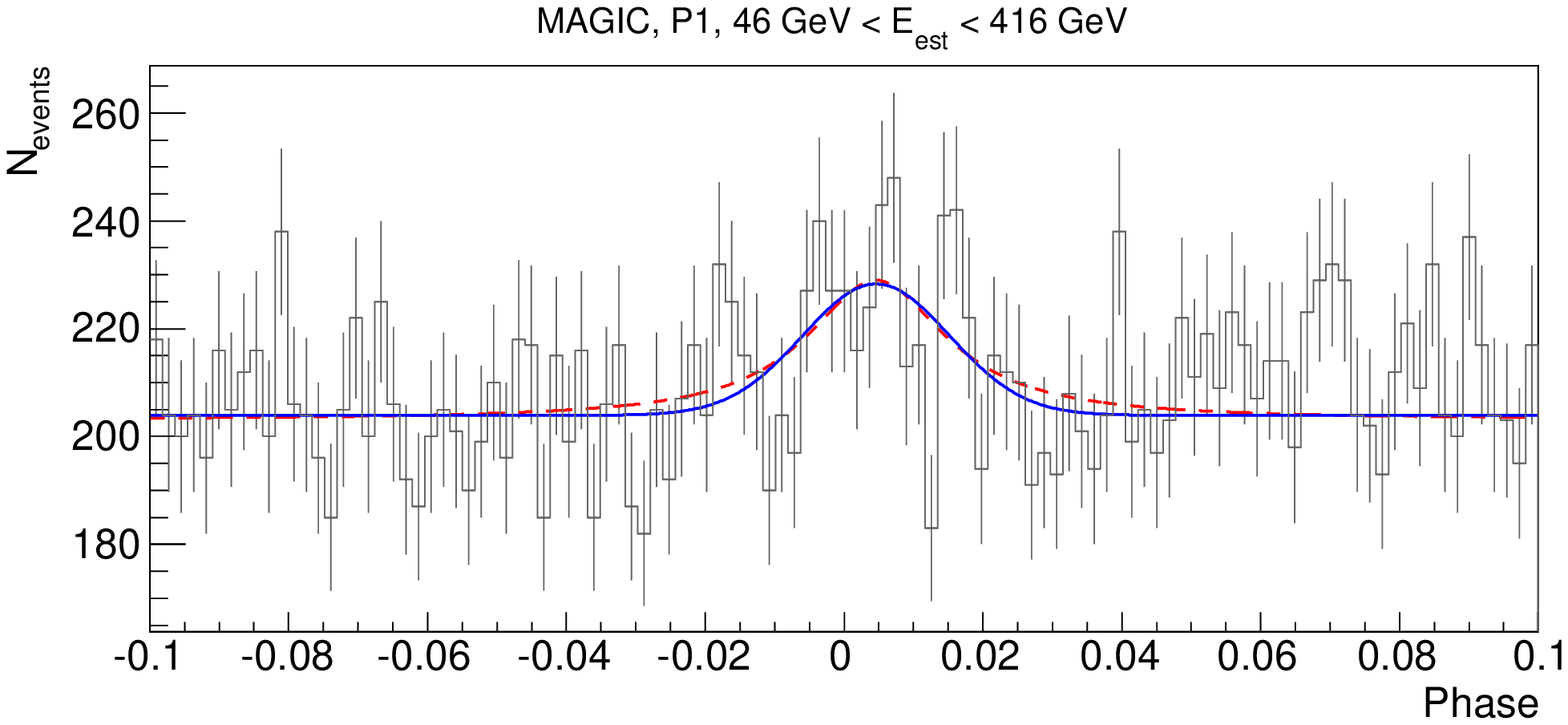}
   \includegraphics[width=8.5cm]{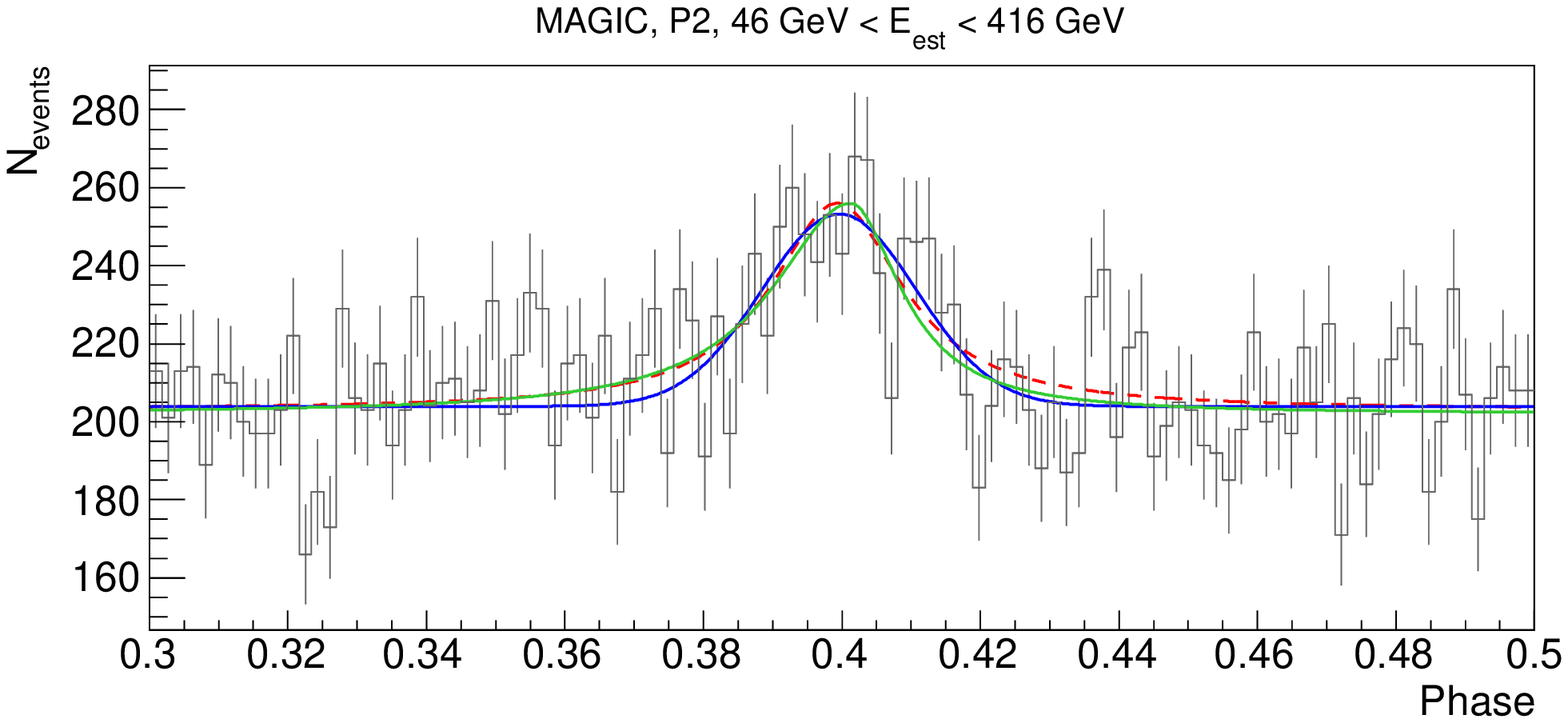}
      \caption{\bref{Close-up display of the two fitted peaks P1 and P2 using a
finer binning and a smaller range. The blue solid curves represent the Gaussian
functions that we use to define the signal phase intervals. The red
dashed curves
are the Lorentzian functions, which allow for wider tails, and the green
solid curve is an
asymmetric Lorentzian function. The latter did not converge for P1 (see
text).}
              }
         \label{FigLCcloseup}
   \end{figure}

\bref{We fitted a very fine-binned all-energy folded light curve, maximizing a
Poissonian
likelihood function that includes two Gaussians or
Lorentzians
over a constant background. \breftwo{The fitted \bref{Gaussian} \breftwo{(Lorentzian)} positions of
P1 and P2 are
$0.005\pm0.003$ \breftwo{($0.005\pm0.002$)} and $0.3996\pm0.0014$
\breftwo{($0.3993\pm0.0015$)},
respectively, with corresponding pulse widths (FWHM) of $0.025\pm0.007$
\breftwo{($0.025\pm0.008$)} and
$0.026\pm0.004$ \breftwo{($0.023\pm0.004$)}.} The signal in P2 is strong
enough to also be fitted with an
asymmetric Lorentzian, which involves more parameters. The results are
displayed in Fig.~\ref{FigLCcloseup}. All fits to our data yield \breftwo{very similar
likelihoods}, which neither supports nor excludes the presence of the tails implied by a Lorentzian
function.
Furthermore, the asymmetric fit does not
yield a significant difference in the leading and the trailing wings of P2.
Hence,
we conclude
that the \breftwo{conservative approach of using a Gaussian parameterization is
sufficient to describe our peaks.}

\breftwo{Notably, there is a positive excess
throughout the region between the two peaks. Most prominently, the
trailing wing TW1 $= [0.04 - 0.14]$ has an excess corresponding to
$3.4\eh{\sigma}$ in the lower-energy bin ($46-138\gev$), which may allow for a
significant detection once more data is collected. A bridge emission
between the peaks in our lowest-energy light curve is also expected if one
considers that in the Fermi-LAT data presented in
\citet{fermicrab}, the bridge emission is evident up to at least $10\gev$, and
it
is denoted as being spectrally harder than the peak emission. However, our
current significance in the bridge region is too low for a spectral
analysis and will not be considered in more detail.
}
}
 
\begin{figure}
   \centering
   \includegraphics[width=9.0cm]{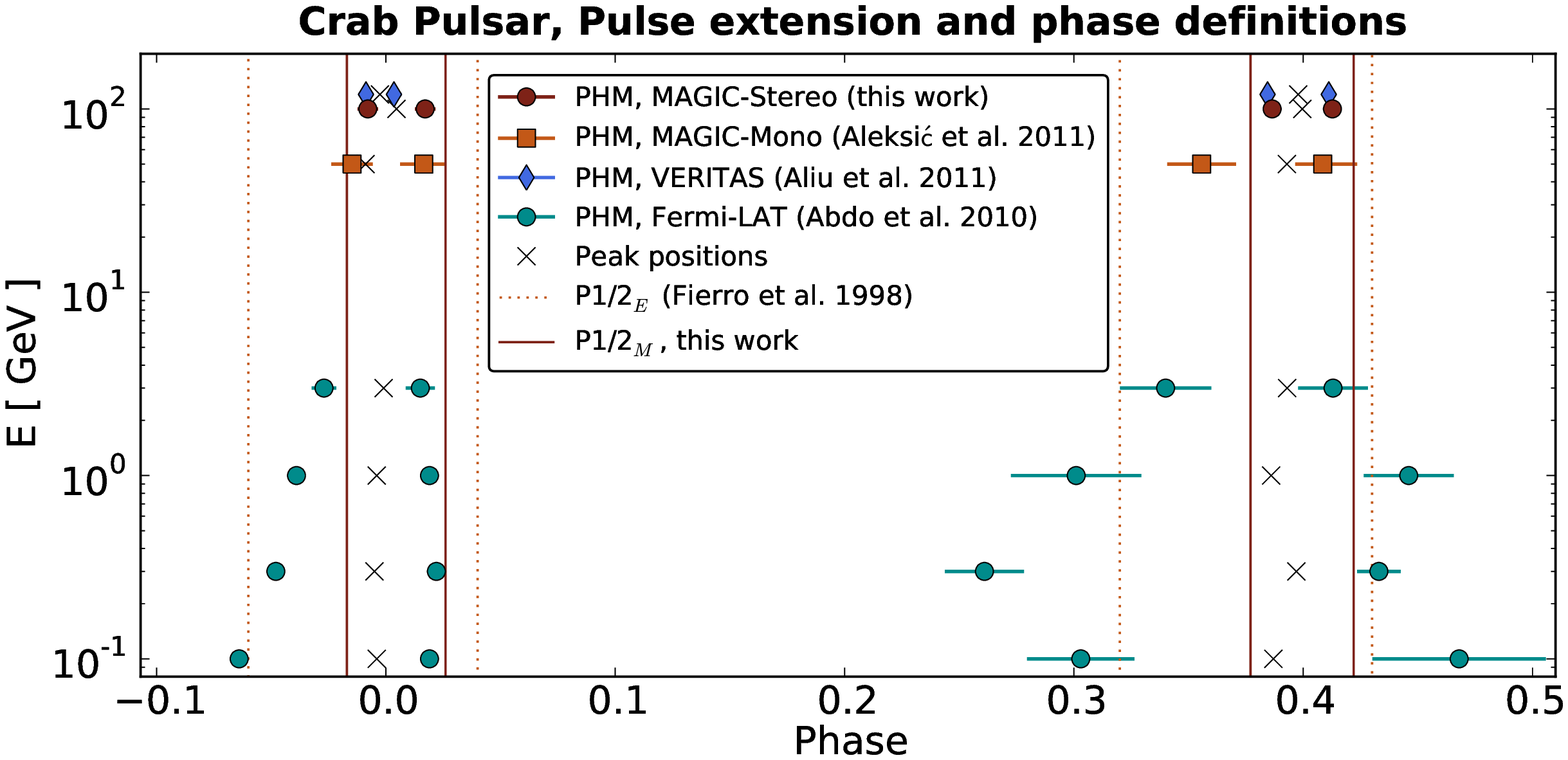}
      \caption{Compilation of pulse profile parameters at different energies,
measured by
Fermi-LAT~\citep[][light circles]{fermicrab},
MAGIC-Mono~\citep[][squares]{magictakacrab}, MAGIC-Stereo
(this work, dark circles)
and VERITAS~\citep[][diamonds]{veritascrab}. The \brefthree{solid
points are the phases of the half-maxima (PHM), while the crosses indicate the
corresponding phases of the peak}. The vertical
lines indicate the phase range
definitions used for the spectra in Fig.~\ref{FigSpectra}.}
         \label{FigPeaks}
   \end{figure}

The peaks \breftwo{we found} are significantly
narrower than those in the GeV regime, and along with MAGIC-Mono and VERITAS data, a consistent trend from GeV to beyond
$100\gev$ can be established (Fig.~\ref{FigPeaks})\footnote{\brefthree{A
correction to the absolute phase values in \citet{fermicrab} was announced on the
Fermi-LAT websites (http://fermi.gsfc.nasa.gov/ssc/data/access/lat/ephems/)
and is incorporated in this plot.}}. Consequently, the
excess we found is much more concentrated than the wide peak ranges \breftwo{defined in
\citet{egretphases} (P1$_\mr{E} = [0.94 - 0.04]$ and
P2$_\mr{E} = [0.32 - 0.43]$}, where \textit{E} stands for EGRET, in contrast to
the \textbf{M}AGIC and
\textbf{V}ERITAS definitions below). Because with too-large phase intervals
one integrates an unnecessarily large number of noise events, we
decided to investigate the signal \breftwo{both in the EGRET intervals and
in narrower, a posteriori defined phase}
intervals, using the Gaussian peak positions $\pm 2
\eh{\sigma}$\bref{, as was done in \citet{veritascrab}}. We obtained
P1$_\mr{M} = [0.983 - 0.026]$ and P2$_\mr{M} = [0.377 - 0.422]$, the excess of which corresponds to $10.4\eh{\sigma}$ after
  \citet[][Eq.~17, see also Table~\ref{table:1}]{lima}. \breftwo{The
low-/high-energy Li~\&Ma significances for P1$_\mr{M}$
(P2$_\mr{M}$) are 4.4/3.3
(7.9/5.9). A listing of the all-energy significances 
can be found in
Table~\ref{table:1}.}

\bref{It is important
to note that the two phase interval definitions are equally valid. The
difference between them is mainly that the wide intervals lead to a higher noise contribution
but are free of any possible selection bias, whereas the narrow intervals have
much lower noise, but are affected by a
minor selection bias.}
\breftwo{The VERITAS results shown in
\citet{veritascrab} were calculated using P1$_\mr{V} = [0.987 - 0.009]$ and
P2$_\mr{V} = [0.375 - 0.421]$, which is still a bit narrower than our
definitions.}

For the emission ratio between the two peaks, we found $0.54\pm0.12$ for P1$_\mr{M}$/P2$_\mr{M}$ and $0.46\pm0.13$ for
P1$_\mr{E}$/P2$_\mr{E}$. 
We also looked for the differences in the pulse shape parameters between the two
energy intervals of Fig.~\ref{FigLC}, but we found no significant changes in
the pulse width, the position, or the relative intensity (for either phase range
definition). \breftwo{This invariance might be related to the fact that
although our
energy range is almost an order of magnitude, the mean energies of the two energy
bins ($80$ vs. $180\gev$) are comparably close to each other; thus, the lever arm is
small compared to the energy-dependent trend in Fig.~\ref{FigPeaks}.}

%

\subsection{Energy spectra}

\begin{figure}
   \centering
   \includegraphics[width=8.0cm]{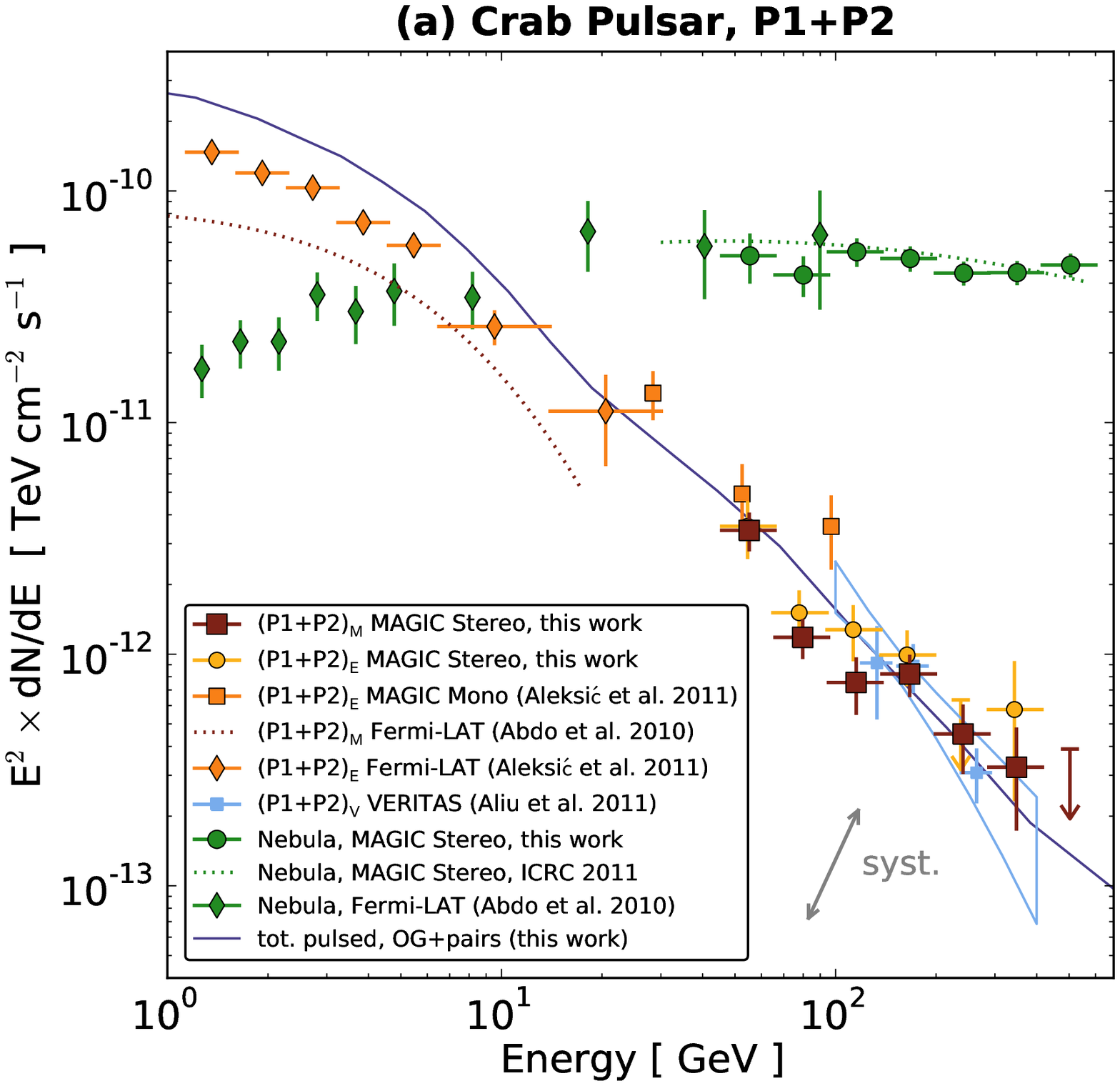}
   \includegraphics[width=8.0cm]{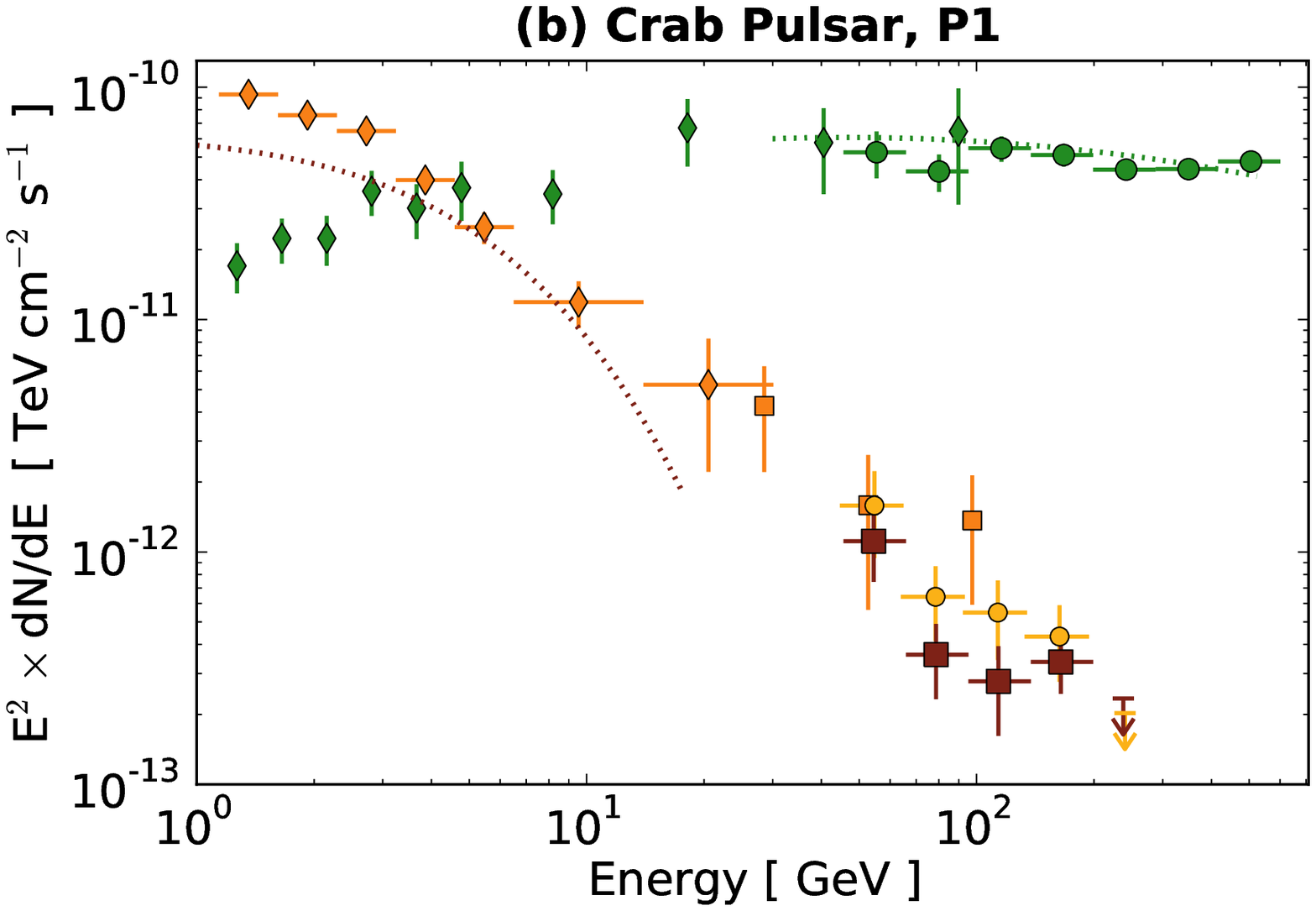}
   \includegraphics[width=8.0cm]{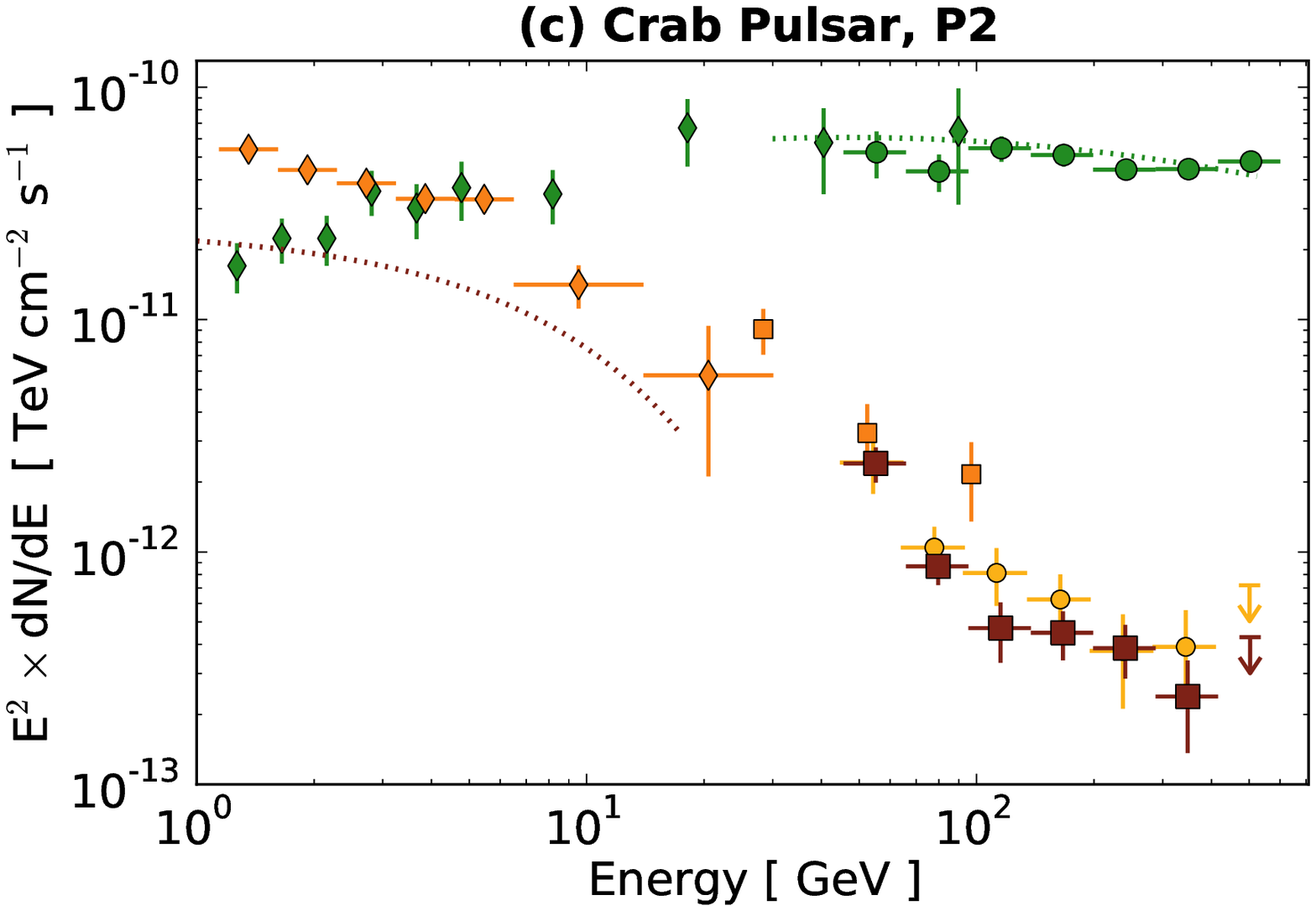}
   \caption{Compilation of the spectral measurements of MAGIC and
Fermi-LAT for the two emission peaks P1 and P2, separately (b, c) and for both
peaks together (a). The VERITAS spectrum is only available (and shown) for
P1$_{\mr{V}}$+P2$_{\mr{V}}$ (light blue squares and solid line). For comparison, the Crab nebula
measurements of MAGIC and Fermi-LAT \bref{(excluding the pulsed component)}
are also shown (green solid circles and diamonds, resp.). Points of similar color refer
to the same phase intervals (dark red for \textbf{M}AGIC, yellow
for \textbf{E}GRET, blue for \textbf{V}ERITAS intervals, and green points for
the nebula spectrum, see also text). The blueish solid line in the
upper plot is the \bref{model} discussed in Sect.~\ref{discussion}
\bref{and it is above
the points because it includes the bridge emission.
The displayed systematic error of the
MAGIC-Stereo measurement corresponds to a shift of $\pm17\pct$ in E and $\pm19\pct$ in flux.}
}
              \label{FigSpectra}%
    \end{figure}

We calculated the energy spectra for (P1+P2)$_\mr{M}$, P1$_\mr{M}$ and
P2$_\mr{M}$, which are shown as the red squares in  Fig.~\ref{FigSpectra}, and
for comparison, we also calculated the spectra for the unbiased
\breftwo{EGRET} intervals (see above), which are shown as the yellow circles.
The latter \breftwo{can be compared directly to previous studies, including
the monoscopic MAGIC observations}. Given that \breftwo{the EGRET intervals}
cover 21\pcnt\ of the
whole phase, they cause a higher background noise than the MAGIC
phase ranges, which cover only 8.8\pcnt. \breftwo{The VERITAS phase intervals cover 6.8\pcnt\ of the whole phase,
which is three times less than the EGRET definitions.
Although most of their narrow pulse may indeed be contained in this interval,
one may expect a certain discrepancy in flux related to this difference in selection.}

\bref{The spectra we obtain for \breftwo{the EGRET intervals} are compatible
with the monoscopic measurements from~\citet{magictakacrab}, considering that the statistical deviations are
at most $\sim 2\eh{\sigma}$ and many of the systematic errors of the two measurements are
independent. 
\breftwo{Our stereoscopic measurements, however, support the possibility that the
gamma-ray energy of MAGIC-mono data may have been over-estimated, as already discussed
in \citet{takathesis}.}
}

\breftwo{The EGRET and MAGIC phase definitions do not result in significantly different
P1 and P2 spectra,} although the points of the latter are systematically
somewhat below the former. This is self-consistent, because \breftwo{the EGRET intervals}
enclose \breftwo{the MAGIC intervals} and shows that the selection
\breftwo{bias that affects the latter} is probably very small.

To
determine the spectral parameters \breftwo{and $\chi^2$ values}, we applied a forward unfolding \citep{unfolding},
which is the most robust method to parameterize the data.
The \breftwo{spectra} could be
described \breftwo{by} power laws as shown in
Table~\ref{table:1}. \bref{It should be noted that the $\chi^2$ values
that we found are not optimal, especially for the spectrum of P1$_{\mr{M}}$. However, the
significance of this inconsistency ($2.6\eh{\sigma}$ pre-trial) is too low to claim a feature
with the data we present here, especially if the systematic
uncertainties are considered.} 

The ratio of the normalization constants between P1 and P2 at $100\gev$ is
$0.4\pm0.2$, which is consistent with the values directly derived from the light curves.
We cross-checked the (P1+P2)$_\mr{M}$ spectrum by comparing the
2009/10 data to 2010/11 data, on- to wobble-mode data, two zenith angle ranges,
two quality cut levels and four unfolding algorithms, and found that the
spectrum was stable
within the
errors.

\breftwo{To ensure a good understanding of all possible systematic
effects, we furthermore
determined the Crab nebula spectrum from the data taken in wobble mode,
analyzing the same energy range with the same
energy binning. The \bref{n}ebula spectrum that we obtained with our cuts (see
Fig.~\ref{FigSpectra}) agrees with both the recent
Crab nebula analysis in \citet{icrccrabnebula} and the Fermi-LAT data in
\citet{fermicrab}, which confirms the good performance
of our spectral analysis down to $46\gev$. 
\bref{Notably, also the Crab nebula flux of the lowest-energy point, which is at approximately
$55\gev$, agrees within errors with
the function derived with a combined Fermi-LAT/MAGIC fit in~\citet{icrccrabnebula}. This fit function is basically independent of
the lowest-energy MAGIC point because it is determined by the statistically much more precise points at higher and
lower energies. From these results,} we find no indication that the total \breftwo{systematic}
flux uncertainty is beyond the
standard low-energy numbers given
in ~\citet{magicstereoperformance}\footnote{This argument may be regarded as a calibration of the pulsar spectra on the nebula
spectrum, a method that is not applicable on the nebula spectrum
itself, but holds for any other source.}. These \breftwo{systematic}
uncertainties are $17\pct$ on the energy scale and $19\pct$ on the flux
normalization, which is displayed in the upper panel of Fig.~\ref{FigSpectra}. Assuming a photon index of 3.6, the total flux
uncertainty \textit{including} a possible energy bias is therefore
$\sim44\pct$ at low energies.} The uncertainty of the spectral index of such a soft
spectrum is approximately $0.2$. 
\breftwo{All MAGIC spectra shown in Fig.~\ref{FigSpectra} are unfolded; thus,
the statistical errors
are correlated by $20-60\pct$, reflecting our energy resolution
and bias, which vary from $15-40\pct$, depending on the energy \citep[see][]{magicstereoperformance}.}

\brefthree{Figure~\ref{FigSpectra} also shows the Fermi-LAT spectra in the
EGRET intervals as determined in~\citet{magictakacrab}. They extrapolate
consistently to the monoscopic and stereoscopic
spectra within systematic and statistical uncertainties.}
\breftwo{To estimate a Fermi-LAT spectrum for the MAGIC phase range definition, we
summed up the matching phase-resolved fit functions
provided in \citet{fermicrab}. Because their flux constants are differential in
phase, the emission from the partly covered phase intervals could also be
approximated. We find that our narrower intervals lead
to substantially lower GeV equivalent flux spectra.}

In general, when comparing our energy spectra to those extracted from Fermi-LAT, VERITAS or
MAGIC-Mono data, it is important to bear in mind that in addition to the
different phase interval definitions, all of these experiments
suffer different and energy-dependent systematic uncertainties that may lead
to discrepancies on the order of $10-30\pct$ in
energy.

\begin{table}
\caption{Results of the spectral fits.
}             
\label{table:1}      
\begin{tabular}{c c c c c c}        
Phase & $S\tin{det}$\tablefootmark{a} & $f_{100\gev}$\tablefootmark{b} &
Phot. index & $\chi^2/\mr{n.d.f.}$\tablefootmark{c} & prob.\tablefootmark{d}\\    
\hline                        
   (P1+P2)$_{\mr{M}}$ & 10.4 & $13.0\pm1.6$ & $3.57\pm0.27$ & $10.3/4$ & $0.04$  \\      
   P1$_{\mr{M}}$               & 5.5  &  $3.9\pm1.7$ & $4.0\pm0.8$   & $9.3/2$ & $0.01$ \\      
   P2$_{\mr{M}}$               & 9.9  &  $8.8\pm1.0$ & $3.42\pm0.26$ & $6.1/5$ & $0.30$ \\      
\hline                                   
   (P1+P2)$_{\mr{E}}$ & 7.7  & $15.5\pm2.9$ & $3.9\pm0.4$   & $9.5/4$          & $0.05$ \\      
   P1$_{\mr{E}}$               & 3.9  &  $6.5\pm2.0$ & $3.3\pm1.0$   & $3.8/2$ & $0.15$ \\      
   P2$_{\mr{E}}$               & 8.0  & $11.2\pm1.9$ & $3.7\pm0.4$   & $7.2/5$ & $0.21$ \\      
\hline                                   
\end{tabular}
\tablefoottext{a}{Detection significance after \citet[][Eq.~17]{lima}}\\
\tablefoottext{b}{Flux at $100\gev$ in units of
$10^{-11}\eh{cm^{-2}\,s^{-1}\,TeV^{-1}}$}\\
\tablefoottext{c}{Number of degrees of freedom taken from the distribution of
estimated energies, which may deviate from the number of unfolded points in
Fig.~\ref{FigSpectra}.}\\
\tablefoottext{d}{The fit probabilities calculated from the $\chi^2$ values do not include systematic effects.}
\end{table}

\section{Discussion and conclusions}\label{discussion}

We found a pulsed VHE gamma-ray signal from the Crab pulsar that allows us to present
spectra with an unprecedentedly broad energy range and phase resolution. For
completeness and comparison, we provide analyses for both the previously
used phase intervals in \citet{egretphases}, and the narrower peaks that we
find in our folded light curves. The range of our spectra is about
one order of magnitude, and, along with the MAGIC-Mono~\citep{magictakacrab}
and the Fermi-LAT data~\citep{fermicrab}, comprise the first gamma-ray
spectrum of the Crab pulsar from $100\eh{MeV}$ to $400\gev$ without any gap. 
On the high-energy end, this result agrees with the recently published
VERITAS spectrum of P1+P2 above $100\gev$, including also the positions and
the remarkably narrow widths of the two pulses. 

\bref{
To interpret the observed pulsed spectrum in the context of the
outer-gap (OG) model,
we follow the same method as described in Sect.~8.2 of \citet{magictakacrab}. In
this framework, the \breftwo{VHE compontent of the spectrum is the} inverse Compton radiation
of secondary and tertiary electron pairs on magnetospheric IR-UV photons. 
To derive the expected gamma-ray flux of this scenario, we solve the set of
Poisson equations for the non-corotational 
potential (Eq.~[9] in \citet{magictakacrab}) with the Boltzmann equations for the created 
electrons and positrons and the radiative transfer equation of the emitted photons.
}

\bref{
We present our theoretical calculation
of the total pulsed spectrum as a violet solid curve in the upper plot of
Fig.~\ref{FigSpectra}. In this calculation, the angle between the 
rotational and the magnetic axes is assumed to be $\alpha=65^\circ$, and the observer's
viewing angle is $\zeta=106^\circ$.

\breftwo{
In \citet{magictakacrab},
the calculations of both $E_\parallel$ 
(the electric field component projected along the local magnetic field line,
which accelerates $e^{\pm}$)
and the resultant primary gamma-ray emissions (curvature+IC) were carried out within
$0.7\,R_{\mathrm{LC}}$ from the rotation axis, 
where $R_{\mathrm{LC}}$ is the radius of the light cylinder.
In our new calculation, to take account the strong primary IC emission 
that becomes important near the light cylinder,
we extend the calculation region up to $0.9\,R_{\mathrm{LC}}$ for $E_\parallel$ 
and up to $1.5\,R_{\mathrm{LC}}$ for primary gamma-ray emissions,
after confirming that the emission above $1.5\,R_{\mathrm{LC}}$ is negligible.
Here, $0.9\,R_{\mathrm{LC}}$ is a safe upper boundary for the $E_\parallel$
calculation, because $E_\parallel$ is anyway diminished at 
$0.9\,R_{\mathrm{LC}}$ owing to the
curving-up field-line geometry towards the rotation axis
near the light cylinder.
}

A remarkable consequence of this extended calculation is an increased inward flux of primary gamma
rays originating from the upper side of the gap, which leads to a higher abundance of pair-produced $e^{\pm}$ at lower
altitudes ($<0.6\,R\tin{LC}$).
\breftwo{This screens the original $E_\parallel$ and hence} reduces the
curvature-radiation component in the primary spectrum. This reduction makes our new
calculation more compatible with the Fermi-LAT data at GeV energies but
\brefthree{does} not
significantly affect the secondary and tertiary components at energies
beyond a few tens of GeV.
}
\bref{
Hence, we conclude that our revised model can reproduce
the total pulsed spectrum between $1$ and $400\gev$ well (see also
\citet{nepomukcrabtheory} for an analytical argument of this
process).
}

\bref{
However, a remaining caveat of our new calculation is that it still includes
the bridge
emission that is not contained in the spectra of only P1+P2. Therefore, it is
still above the Fermi-LAT flux points in
Fig.~\ref{FigSpectra}. A phase-resolved modeling is ongoing and will be
presented in the future. In general, it is however difficult to compute the
 spectral shape above 100~GeV with high precision in the present OG model
 for the Crab pulsar. This is because the photon-photon cross section, and
therefore the gamma-ray absorption, depends on the square of the collision angle, which
is typically a few degrees. Hence a small variation in the geometry can have
a large impact on the flux that escapes the pulsar. Thus, our model should not
be interpreted as a hard
quantatative prediction; instead, it is meant to show that the hard component
we see in the experiment can quantitatively be met within the present understanding
of the OG model. Similarly, the slight modulations of the power law component are not to be
interpreted as a significantly predicted feature.
}

Other possible Ans{\"a}tze \bref{to explain the VHE emission} include the production of inverse Compton radiation in
the unshocked pulsar wind outside the light cylinder by pulsed photons
\citep{felixwind, aharonianpulsarwind}, a striped pulsar wind
\citep{stripedwind}, or the annular gap model presented in \citet{annulargap}. The two
crucial spectral features to establish to test these models
are the expected spectral upward-kink in the transition region between the
curvature and the hard
component, and the detection or exclusion of a terminal cutoff at a few hundred
GeV.

\breftwo{Another topic that we will be able to address with a $2-3$ times larger
dataset is the energy dependence of the pulse shape parameters. The narrowness
of the pulses and its evolution with energy are a stringent requirement that
the theoretical modeling must fulfill because the folded light curve is
almost stable against systematic
uncertainties. Moreover, the indication of pulsed emission in the trailing wing
of P1 may indicate that a VHE signal between the two peaks might be within reach
for low-threshold IACT systems.} The MAGIC telescopes, which are being upgraded
in 2011/12, can \bref{address these topics} in the coming years when more
data will improve the statistical precision of the measurements.

\begin{acknowledgements}
We would like to thank the Instituto de Astrof\'{\i}sica de
Canarias for the excellent working conditions at the
Observatorio del Roque de los Muchachos in La Palma.
The support of the German BMBF and MPG, the Italian INFN, 
the Swiss National Fund SNF, and the Spanish MICINN is 
gratefully acknowledged. This work was also supported by 
the Marie Curie programme, the CPAN CSD2007-00042 and MultiDark
CSD2009-00064 projects of the Spanish Consolider-Ingenio 2010
programme, grant DO02-353 of the Bulgarian NSF, grant 127740 of 
the Academy of Finland, the YIP of the Helmholtz Gemeinschaft, 
the DFG Cluster of Excellence ``Origin and Structure of the 
Universe'', the DFG Collaborative Research Centers SFB823/C4 and SFB876/C3,
\bref{the Polish MNiSzW grant 745/N-HESS-MAGIC/2010/0 and the Formosa Program between
the National Science Council in Taiwan and
the Consejo Superior de Investigaciones Cientificas in Spain
administered through the grant number NSC100-2923-M-007-001-MY3.}
\end{acknowledgements}

\bibliographystyle{aa}
\bibliography{aa}

\end{document}